\documentclass{nature}
\usepackage[format=plain,font=bf,justification=justified,singlelinecheck=false]{caption}

\title{Galaxies as fluctuations in cosmic stellar liquid}

\author{Alexander B. Kashuba}

\begin{document}

\maketitle

\begin{abstract}
Large self-gravitating stellar systems share with correlated liquids \cite{Kadanoff} in condensed matter physics a pattern of hierarchical density variations. While it takes the microscopic time resolution to discern the correlated dynamics of the critical opalescence \cite{Kadanoff}, characteristic astronomical times hide fluctuational dynamics of stellar liquids, where, governed by interstellar correlations, denser clusters of stars assemble and disperse. For a semi-isolated galaxy, these dynamical fluctuations are dense globular clusters \cite{Harris,GC,BS}. For a galaxy cluster \cite{Abel,Xray,GL}, these dynamical fluctuations are the member galaxies, elliptical ones in the interior. Bright over-density fluctuations, galaxies, are exhibits of only a small fraction of stars found in a cosmic stellar liquid, the dark matter \cite{Zwicky,Rubin}. Here I report a fluctuational gravitational collapse as a property of a self-gravitating system in the virial equilibrium.
\end{abstract}

Galactic structures have been a subject of intense study and competing concepts \cite{BT}. The stellar dynamics \cite{BT}, as developed over the last century, averages out the gravitational pull of many billions of constituent stars and describes it by a smooth gravitational potential $\Phi(\vec{r})$. The established approximation is that trajectories of stars in this potential are uncorrelated with each other over long  times. Rare passing events, collisions, randomize stars motions after many orbits completed but do not induce long range correlations until the Spitzer time \cite{Spitzer} relaxation sets in. This is basically the kinetic theory of gases \cite{Chavanis}. Here I prove this concept wrong. Ergodicity and the Poisson statistics will be keys. In the kinetic theory, a gas is completely described by an one particle distribution function $f(\vec{r},\vec{v})$, let it be static. Three fundamental laws constrain it. First, the dynamical equation associated with the Liouville's theorem determines the distribution function, $f(\vec{r},\vec{v})$, for a given gravitational potential $\Phi(\vec{r})$. Second, the Poisson's equation determines $\Phi(\vec{r})$ as originating from the mass source, thus, closing up the self-consistency. Third, the global virial theorem holds. For the central symmetry, the Jeans' theorem requires $f(\epsilon,\vec{l}^2)$ to depend on two integrals of motion: the energy and the angular momentum vector squared. These constraints alone are not restrictive enough and the Eddington's formula \cite{BT} provides the isotropic kinematics for any distribution of the mass inside a galaxy which is, thus, possible. One practical spin-off of this belief in richness of galactic structures is the galactic archeology which reconstructs old merger events by studying the present time $f(\vec{r},\vec{v})$ \cite{archeol}. The Liouville's theorem conserves the phase space. The probability to find a star or many stars in some small part of the phase space obeys the Poisson statistics, the defining property of an ideal classical gas. Given a particle distribution function, satisfying all the above conditions, together with a representative finite system and a definition of an event in the phase space, the Newton evolution, dynamics and gravity, of this system should reveal the Poisson statistics of events as time independent until the Spitzer time.

A choice of the initial particle distribution function matters. In statistical physics any globally conserved property, such as the ratio of the total numbers of atoms A and B in a binary mixture, is satisfied also locally, neglecting small fluctuations. In the thermal equilibrium the total energy of gas is divided equally between molecules, each serving as a small subsystem. In the case of the virial equilibrium, the global virial theorem condition can be divided equally between local subsystems in the form of a virial equipartition:
\begin{equation}
\langle \vec{v}^2(\vec{r})\rangle =\frac{1}{2} \Phi(\vec{r}).
\label{equi}
\end{equation}
It strongly constraints the galactic structure (see two remarks in Methods). Possible are centrally symmetric structures parameterized by one dimensionless parameter $0<\Theta<\infty$, like the temperature \cite{hazy}. The gravitational potential (inverted) reads:
\begin{equation}
\Phi(\vec{r})=2\sigma\left(\frac{1}{1+\vec{r}^2/a^2}\right)^{\frac{2\Theta}{1+4\Theta}},
\label{Potential}
\end{equation}
where $a$ is the core radius, $\sigma=\langle \vec{v}^2 \rangle_c$ is the central velocity dispersion. The mass distribution function reads:
\begin{equation}
f(\vec{r},\vec{v})= \frac{4\Theta}{(1+4\Theta)^2 G a^2} \left[ \frac{1}{\sqrt{2\Phi(\vec{r})-\vec{v}^2- \frac{1}{a^2} [\vec{r}\times\vec{v}]^2}} + \frac{\sqrt{\pi}\Theta}{\sqrt{\sigma}} \frac{\Gamma\left(7+\frac{1}{\Theta}\right)}{\Gamma\left(\frac{9}{2}+\frac{1}{\Theta}\right)}  \left(\frac{2\Phi(\vec{r})-\vec{v}^2}{4\sigma}\right)^{\frac{7}{2}+ \frac{1}{\Theta}}\right]
\label{PDF}
\end{equation}
where $G$ is the Newton gravitation constant. This system is not gravitationally bound and has an infinite extent. Usually, self-gravitating systems are modeled on the Solar System as having the finite total mass, equal to one in the Henon units. The most concentrated mass in equation (\ref{PDF}), at large $\Theta$, is described by the Plummer model \cite{GC,BT} in the interior, but has a thin divergent mass tail at $r>\sqrt{12\Theta} a$. Recently, such faint extended tails have been observed in three globular clusters \cite{Kuzma}. Apparently, elliptical galaxies are spherical in the interior \cite{KFCB,VEGAS}. Coreless galaxies have low $\Theta$ whereas extra light galaxies have higher $\Theta$ \cite{KFCB}. With an additional, gas extinction, parameter the photometry of galaxies \cite{KFCB,VEGAS} can be matched with equation (\ref{PDF}) in the interior, revealing $\Theta\sim 0.1$. Kinematic survey of galaxies \cite{SLUGGS}, if interpreted with the equation (\ref{PDF}), requires no dark matter up to the largest probed distances and also gives $\Theta\sim 0.1$. The equation (\ref{PDF}) describes progressively more radial motion of stars in the outskirts as has been confirmed in the numerical simulations \cite{Dekel,Heggie}.

We generate randomly and independently $N=99999$ particles, points in the phase space $(\vec{r}_i,\vec{v}_i)$, using equation (\ref{PDF}) at $\Theta=1/10$ and $a=6\sqrt{6/7}$ inside a sphere of radius $R=80$ (see Methods). All particles have the unit mass. The mass outside the sphere is assumed to be smeared spatially and its gravity is neglected. The central velocity dispersion $\sqrt{\sigma}=68.652$ ensures the exact number $N$ of particles. A spherical mirror at $R=80$ reflects elastically all outgoing particles. The total energy and the angular momentum inside are conserved. The center of mass and the total momentum fluctuate due to the pressure applied by the mirror. This system evolves according to the Newton laws of dynamics and gravitation, $G=1$, see Methods. Typical orbit period is $1/\sqrt{G\langle\rho\rangle}=4.6$ in our units. The particle distribution function (equation (\ref{PDF})) is dynamically conserved in the limit $N\to\infty$.

On time tics $\Delta t=0.01$ new positions of all particles in the phase space are examined. Groups of particles compact in the phase space (a substitute for the gravitationally bound cluster) are searched for. Until the Spitzer time, estimated to be 3000 in our units, no close binaries will form \cite{Heggie}. Unlike the condensed matter physics where correlations are often two-particle in nature, in gravitationally correlated stellar liquid, fluctuating groups should contain many particles. Too many will amplify sensitivity to a minor change of $f(\vec{r},\vec{v})$. Nine is our choice. Methods describes in detail our phase space defined trigger. On each tic the trigger counts few compact groups in the gas, events, and this number is expected to obey the time independent Poisson statistics.

However, Figure 1 asserts to the contrary. Numerated by letters A,B,C,D,E,F,G,H if comprising totally different particles and numbers C1,C2,...C64, D1 if having some particles in common (i.e. overlapping in the phase space), 73 distinct compact groups of particles are shown in Figure 1 versus the time they have triggered the event counts, 473 in total. The most rich multi-group C involves in total 16 particles aggregating in groups of 9,10 and even 11 particles (the group C33). Some particles are anchoring many events some are just passing by. All this happens as this multi-group falls from $R=60$ to $R=19$ accelerating from $V_r=-59$ to $V_r=-87$. The multi-group D involves 10 particles in total and is bounded outside $V_r=12$ at $R=22$. The early, left half of Figure 1 is different visually and statistically (few events per tic) from the later, right half (twenty events per tic). The distribution of 73 groups in time proves the evolutionary change with the statistical confidence $5\sigma$.

\begin{figure}
\includegraphics{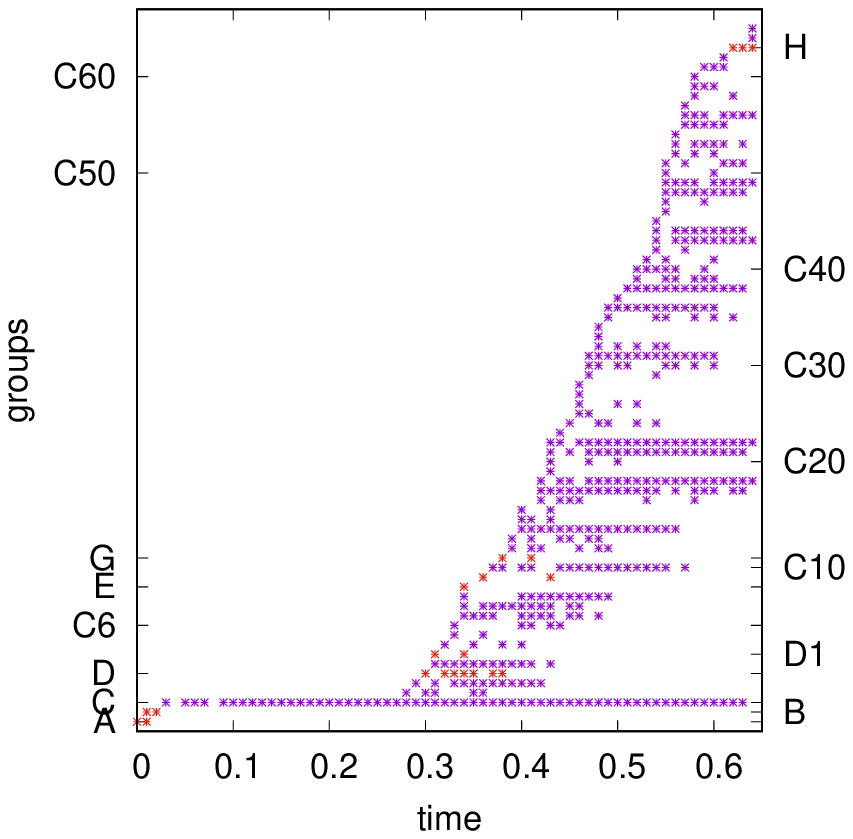}
\caption{Trigger's event counts versus time. Different groups of particles are sorted out over the vertical axis.} \label{Evolution}
\end{figure}

\vspace{16cm}

Obviously, we see in Figure 1 some sort of the gravitational collapse. The question is why in seemingly "hot" gas in the virial (equation (\ref{equi})) not thermal equilibrium, it is developing so fast. In the outskirts the first term in equation (\ref{PDF}) dominates. Velocities here are almost radial. Particles here are crowded towards the two speed limits $\pm\sqrt{2\Phi(\vec{r})}$ (the escape velocity) in and out, like cars on a highway. In the reference frame of one crowd the "cold" Jeans instability, $\tau\sim 1/\sqrt{G\rho}$, can develop. However, the opposite "hot" crowd can scatter it. The virial equilibrium is, thus, a balance of a cold-hot duality. Figure 1 reveals a fluctuational gravitational collapse on a smaller scale. Inside a compressed cluster many particles are found in almost the same point, i.e. strongly correlated. They exert a combine long range gravitational force from there. The number of particles collapsing into compact clusters is exactly how much the Spitzer time underestimates evolutionary changes. In the end of the time evolution (Figure 1) interparticle correlations will presumably grow up to maturity. 

On each time tic in numerical simulations we see few compact groups (Figure 1). On the present tic of astronomical time, globular clusters \cite{Harris,GC} are observed in the outskirts of all galaxies studied \cite{BS}. What if they are fluctuations too. Distinct populations of stars, found in globular clusters \cite{BS,biMod}, suggests so. Evidence for the young age of some globular clusters has emerged \cite{youngGC}. Historically, the first model for a globular cluster was the Plummer model \cite{GC,BT}. Equation (\ref{PDF}) approximates it for $\Theta>1$. Comparing that with the galactic values we conclude that the over-density fluctuations have much larger $\Theta$ than the host. Other parameters $a$ and $\sigma$ scale down from the host to fluctuations. In the dense host, fluctuations could grow extremely dense \cite{dense}. Postponing the light to mass discrepancy until later, let us describe a galaxy cluster in this framework. Thus, here we expect $\Theta\sim 0.01$. Gravitational lensing observations \cite{GL} provide $a\sim 50\ kpc$ and $\sqrt{\sigma} \sim 2000\ km/s$, indeed much larger than that for the galaxies. The core mass density is low, $0.01\ M_\odot/pc^3$, a mere background for over-density fluctuations. An equation associated with the virial equipartition theory \cite{hazy}, but accounting for the cosmic expansion,:
\begin{equation}
\vec{\nabla}^2 \Phi + \Phi^{5+\frac{1}{\Theta}} +\frac{1}{4 \Theta} \frac{\left( \vec{\nabla} \Phi \right)^2}{\Phi}+\frac{3}{2}H^2=0,
\label{Hubble}
\end{equation}
where $H$ is the Hubble constant, gives the radius of a typical isolated galaxy cluster $R\sim 10\ Mpc$ and the total mass $6\cdot 10^{14}\ M_\odot$. The average density of a galaxy cluster is only three times the density in the outskirts.

Stellar systems consist roughly of equal amount of stars and gas. In the virial equilibrium the particle distribution function (equation (\ref{PDF})) applies to both. Molecules move as fast as stars. They emit, in collisions, hydrogen lines and the bremsstrahlung x-rays \cite{Xray}. The gas is admixed in the fluctuational gravitational collapse (Figure 1). However in addition, a star can hold a dense circumstellar globule of gas enshrouding it \cite{glob}. The equation (\ref{PDF}) for $\Theta\to \infty$ applies here, with the core and the outskirts representing a star and a gas. By equating the luminosity of the star and the black body radiation emitted by the globule's atmosphere, the globule radius is inferred:
\begin{equation}
r_g=\sqrt{\frac{4\pi\sigma_{SB}}{L}}\left(\frac{GMm}{6k_B}\right)^2,
\end{equation}
where $\sigma_{SB}$ is the Stefan-Boltzmann constant, $k_B$ is the Boltzmann constant, $m$ is the hydrogen atom mass, $L$ and $M$ are the luminosity and the mass of a star. The ratio of the mass of the globule to the mass of the host star is $\sim 1/\Theta$. Dust condensation, possible in the hydrogen gas at low temperatures $T<35\ K$ \cite{dust}, makes the globule dark.  Because globules are small the Universe is transparent. The temperature of the globule's atmosphere is:
\begin{equation}
T_g=\sqrt{\frac{L}{4\pi\sigma_{SB}}}\frac{6k_B}{GMm}.
\end{equation}
Massive main sequence stars are too luminous for the dust to ever gather, whereas $T_g$ for light main sequence stars bumps into the temperature of the cosmic microwave background. In the galactic interiors the ram pressure of the impacting gas disperses the globule, $\rho v^2 \sim k_B T_g M/r^3_g m$, when a star, falling from the outskirts, reaches a threshold dense environment. Cold gas release results in the gas density discontinuity, the cold front \cite{CF}. Less dense galaxy clusters remain dark in the centre. On the way out, it takes $3\cdot 10^7$ years for a star to gather the comoving gas into a globule. In the galactic shell $\Delta r\sim 10\ kpc$, where this happens, the gas sneaks in the interiors inside cold globules but outflows ionized. A star can accrete the hydrogen from its globule. Upon the release, it will shift on the Hertzsprung Russell diagram to a cooler and younger line. Small stars never shrouded by a globule or stars staying in the bulge, by chance, remain original \cite{Old}. A globule impacting on the red giant star can strip it from its tenuous atmosphere leaving a blue core star behind. By gathering hydrogen gas it will return along the horizontal branch \cite{GC} to the red giant. Thus, the globular clusters population is bimodal in color \cite{BS,biMod}.

Aside the Big Bang nucleosynthesis, a concept of baryonic cosmic stellar liquid matches the observations. Being in place early on at $z>10$ \cite{Ellis}, when the microwave background heat sublimates the dust, the young cosmic stellar liquid appears to us as the uniform cosmic infrared background. Being very dense and moving fast then, the constituent gas emit hydrogen lines. Gathering of dark globules around stars initiates at $z\sim 10$ \cite{z10}, first for stars in the middle of the main sequence and in the less dense environments. The light amplifies the mass variations. At $8>z>1$, the cosmic stellar liquid dims with a widening gap between visible blue and red stars. Such spectrum was interpreted as an extra star formation activity \cite{SB}. Two colliding globules produce luminous, $L\sim 10^{32}\ W$, event visible in emission lines and x-rays for tens of years. The parameters of over-density fluctuations scale like $(a,R)\sim (1+z)^{-1}$ and $\sigma \sim 1+z$, \cite{zR,hazy} whereas shadows cast by the globules are more prominent at this epoch. Morphologically, visible then galaxies are limited by shadows and irregular. Dark over-density fluctuations then are observed as submillimeter galaxies. In the local Universe, over-density fluctuations in the initial stage of the growth are luminous infrared galaxies \cite{LIG} and in the compressed stages are elliptical galaxies. If hurled into a low density environment, the latter survive by rotation as spiral galaxies.

\subsection{Methods}

For large $\Theta$, the self-gravitating gas is compact and has thin outskirts. The wall here is not a disruption but the gravitational collapse is not expected. For small $\Theta$, the self-gravitating gas has thick outskirts, a favorable condition for fluctuations. However, the wall here is a major disruption. I decided to place it where the half of the virial theorem is satisfied and the total energy equals to zero. Imaginary particles outside the wall contribute the remaining gravitational energy to satisfy the virial theorem in full.

$N=99999$ particles are generated as follows. Six random numbers select uniformly a point in the phase space $|\vec{r}_i|<80$ and $|\vec{v}_i|<138$ as real*8. The term inside parenthesis of equation (\ref{PDF}) is evaluated for $(\vec{r}_i,\vec{v}_i)$ and $\Theta=1/10$, multiplied by a small number $7\cdot 10^{13}$ and then compared to a random positive integer*8 number. If larger a new particle is added, if smaller then repeat. 

Discrete time evolution of the system proceeds in steps of time $dt=4\cdot 10^{-6}$ from the old, superscript $o$, to a new, superscript $n$, configuration:
\begin{equation}
\vec{r}_i^n=\vec{r}_i^o+\frac{1}{2}\left(\vec{v}_i^n+\vec{v}_i^o \right)\ dt
\end{equation}
and
\begin{equation}
\vec{v}_i^n=\vec{v}_i^o+\frac{1}{2}\sum_{j\neq i}\left(\frac{\vec{r}_j^o-\vec{r}_i^o}{\left((\vec{r}_j^o-\vec{r}_i^o)^2+a\right)^{3/2}}+\frac{\vec{r}_j^n-\vec{r}_i^n}{\left((\vec{r}_j^n-\vec{r}_i^n)^2+a\right)^{3/2}}\right) dt
\end{equation}
where $a=10^{-5}$. These equations are iterated three times which gives a convergent new configuration. In comparison, four iterations will not change real*8 space positions of particles a single digit, however, typically two last digits of the real*8 velocities will converge. Systematic errors of the above scheme are $dt^3$ and negligible. Parameters $dt$ and $a$ are commensurate in such a way that a tight collision proceeds smoothly in ten or so steps. The total energy, including $a$, fluctuates around the mean value (two percents of the total kinetic energy) by few $10^{-3}$ due to omnipresent tight collisions in the large system, however, over the time $t=0.7$ the total energy does not diffuse away and stays put. The overall computational effort is $3\cdot 10^{15}$ square roots.

To check if $f(\vec{r},\vec{v})$ is dynamically conserved, I sliced the sphere into 80 shells $i-1<R<i$, $i=1...80$ and have calculated occupation numbers for each shell using equation (\ref{PDF}). The total is $N$. Then, the two totals of the square residue of these and the integer initial and final, at $t=0.7$, particle distributions are $0.91 \cdot 10^5$ and $1.06\cdot 10^5$. This, and probably all, self-gravitating system is slightly contracting initially to facilitate the build up of the interparticle correlations. In the longer run, $t\sim 10$, these oscillations should disappear.

Also, I have measured the diffusion of the particle velocities away from that calculated in the smooth potential $\Phi(\vec{r})$. The observed velocity diffusion is fast, such that after the estimated time $t\sim 100$, shorter than the Spitzer time, particles will forget the initial conditions.

The trigger counts one event if there exists a point in the outskirts $19<|\vec{r}|<76$ and within wide velocity range $|\vec{v}|<109$ such that nine or more particles $(i_1,i_2,...,i_k)$, $k\geq 9$, exist each satisfying the closeness criteria: 
\begin{equation}
\left(\left(\vec{r}_{i_j}-\vec{r}\right)^2+4\right)\left(\left(\vec{v}_{i_j}-\vec{v}\right)^2+14\right)<1250.
\label{trigger}
\end{equation}
The number in the r.h.s. simply adjusts the event rate whereas the pair of numbers 4 and 14 regulate spatial and velocity compactness of the group. Thinking here is simple: if particles are close in velocity then they stay close longer time, if they close in space then the gravitation force is large. Small number 4 selects spatially compact groups whereas large number 14 allows for disperse velocities. Interchanging these two numbers will define another trigger that will produce evolution opposite to and not as dramatic as in Figure 1. The choice of numbers 9 and 9 will result in no noticeable evolution at all. The entire multi-particles phase space seems to evolve during the observation time $t=0.7$. 

After Figure 1 had emerged, I have initiated randomly the same system 23 more times and probed it with the same trigger at zero time without the dynamical evolution. Event counts were none for 12, one for 3, two for 4, three for 1, four for 2 and six for 1 systems, i.e. 1.2 events per tic, consistent with the left half of Figure 1.

Consider a gas composed of species $A,B,C...$. If two identical atoms $A$ and $A$ interact, $m_Am_A/r_{AA}$, symmetry dictates to divide it equally between them. In the case of two different atoms $A$ and $B$, one can otherwise attribute a portion $\frac{1}{2}+\eta_{AB}$ of the interaction, $m_Am_B/r_{AB}$, to the atom $A$. Then, the atom $B$ gets the remaining $\frac{1}{2}-\eta_{AB}$ part. Let $\eta_{AB}$ be antisymmetric. Consider two subsystems containing large numbers of atoms $N_1$ and $N_2$. The fraction of a particular atom in the well mixed gas is the same everywhere $N_{iA}=\nu_A N_i$, where $i=1,2$. Then, the equal part of the interaction is attributed to both subsystems:
\begin{equation}
E_1=\sum_{AB} N_{1A} \left(\frac{1}{2}+\eta_{AB}\right)\frac{m_Am_B}{r_{AB}} N_{2B}= \frac{1}{2} N_1 N_2 \left(\sum_{A}\nu_A m_A\right)^2 \langle \frac{1}{r_{12}} \rangle =E_2,
\end{equation}
irrespective of $\eta_{AB}$. Therefore, the one half in equation (\ref{equi}) is a robust statistical symmetry.

Let the kinematic isotropy condition be imposed in addition to the virial equipartition equation (\ref{equi}). The stress tensor is $T^{ab}=\rho\langle v^a v^b\rangle = \rho \langle \vec{v}^2 \rangle \delta^{ab} /3=\rho\Phi \delta^{ab} /6$. Then, the hydrostatic equilibrium equation:
\begin{equation}
\frac{\partial T^{ab}}{\partial r^b}=\frac{1}{6}\frac{\partial \rho \Phi}{\partial r^a}=\rho \frac{\partial \Phi}{\partial r^a},
\end{equation}
integrates to $\rho \sim \Phi^5$ and the Poisson's equation becomes the Lane-Emden equation of the exponent five. Therefore, the particle distribution function (equation (\ref{PDF})) is unique in the vicinity of the isotropic case. By deforming the potential $\Phi\to \Phi'=(1+\frac{1}{\Theta})^{1/4}\Phi^{1+\frac{1}{\Theta}}$ in the Lane-Emden equation we recover the equation (\ref{Hubble}).

\begin{addendum}
\item[Author Information]
Reprints and permissions information is available at www.nature.com/reprints. The author has no competing financial interests. Correspondence and requests for materials should be addressed to alexander$\_$ kashuba@yahoo.com
\end{addendum}

\begin{thebibliography}{99}
\bibitem{Kadanoff} Kadanoff, L. P., Goetze, W., Hamblen, D., Hecht, R., et al. Static phenomena near critical points: theory and experiment. \textit{Rev. Mod. Phys.} \textbf{39}, 395 (1967).
\bibitem{Harris} Harris, W. E., http://physwww.mcmaster.ca/∼harris/mwgc.dat (2010).
\bibitem{GC} Ashman, K. M. \& Zepf, S. E. \textit{Globular Cluster Systems} (New York: Cambridge University Press, 1998).
\bibitem{BS} Brodie, J. P. \& Strader, J. Extragalactic globular clusters and galaxy formation. \textit{Annu. Rev. Astron. Astrophys.} \textbf{44}, 193 - 267 (2006).
\bibitem{Abel} Abell, G. O., Corwin, H. G. \& Olowin, R. P. A catalog of
rich galaxy clusters. \textit{Astrophys. J. Suppl.} \textbf{70}, 1 - 138 (1989).
\bibitem{Xray} Sarazin, C. L. X-ray emission from galaxy clusters. \textit{Rev. Mod. Phys.} \textbf{58}, 1 (1986)
\bibitem{GL} Wang, X., Hoag, A., Huang, K. -H. et al. Mass reconstruction of the lensing cluster Abell 2744 from frontier field imaging and GLASS spectroscopy. \textit{Astrophys. J.} \textbf{811}, 29 (2015)
\bibitem{Zwicky} Zwicky, F. The redshift of extragalactic nebulae, \textit{Helv. Phys. Acta} \textbf{6}, 110 (1933).
\bibitem{Rubin} Sofue, Y. \& Rubin, V. Rotation Curves of Spiral Galaxies. \textit{Annu. Rev. Astron. Astrophys.} \textbf{39}, 137 - 174 (2001).
\bibitem{BT}  Binney, J. J. \& Tremaine, S. \textit{Galactic Dynamics} (Princeton: Princeton University Press 2008)
\bibitem{Spitzer} Spitzer, L. Jr {\it Dynamical Evolution of Globular Clusters} (Princeton University Press, Princeton 1987)
\bibitem{Chavanis} Chavanis, P. H. Kinetic theory of spatially homogeneous systems with long-range interactions. \textit{Eur. Phys. J.} \textbf{127}, 19 (2012)
\bibitem{archeol} Myeong, G. C., Evans, N. W., Belokurov, V., et al. The Milky Way halo in action space. \textit{Astrophys. J.} \textbf{856}, L26 - 32 (2018)
\bibitem{hazy} Kashuba, A. Statistical mechanics of self-gravitating gas like galaxy. \textit{arxiv}:1702.05429 
\bibitem{Kuzma} Kuzma, P. B., Da Costa, G. S., Mackey, A. D. \& Roderick T. A. The Outer Envelopes of Globular Clusters. \textit{Mon. Not. Roy. Astron. Soc.} \textbf{461}, 3639 - 3652 (2016)
\bibitem{KFCB} Kormendy, J., Fisher, D. B., Cornell, M. E. \& Bender, R. Structure and formation of elliptical and spheroidal galaxies. \textit{Astrophys. J. Suppl.}, \textbf{182}, 216 (2009).
\bibitem{VEGAS} Capaccioli, M., Spavone, M., Grado, A. et al. A VST early-type galaxy survey. \textit{Astron. Astrophys.} \textbf{581}, A10 (2015).
\bibitem{SLUGGS} Brodie, J. P., Romanowsky, A. J., Strader, J. et al. The Sages legacy unifying globulars and galaxies survey. \textit{Astrophys. J.} \textbf{796}, 52 (2014)
\bibitem{Dekel} Dekel, A., Stoehr, F., Mamon, G. A. et al. Lost \& found dark matter in elliptical galaxies. \textit{Nature}, \textbf{437}, 707-709 (2005).
\bibitem{Heggie} Baumgardt, H., Hut, P. \& Heggie D. C. Long-term evolution of isolated N-body systems. \textit{Mon. Not. Roy. Astron. Soc.} \textbf{336}, 1069 (2002)
\bibitem{biMod} Yoon, S., Yi, S. K., and Lee, Y. Explaining the color distributions of globular cluster systems in elliptical galaxies. \textit{Science} \textbf{311}, 1129 -1130 (2006).
\bibitem{youngGC} Hansen, B. M. S., Kalirai, J. S., Anderson, J. et al. An age ifference of 2 Gyr between a metal-rich and a metal-poor globular cluster. \textit{Nature} \textbf{500}, 51 - 54 (2013).
\bibitem{dense} Strader, J., Seth, A. C., Forbes, D. A. et al. The densest galaxy. \textit{Astrophys. J.} \textbf{775}, L6 (2013).
\bibitem{glob} Shinnaga, H., Phillips, T. G., Furuya, R. S. \& Kitamura, Y. Warm extended dense gas lurking at the heart of a cold collapsing dense core. \textit{Astrophys. J.} \textbf{706}, L226-229 (2009).
\bibitem{dust} Yang, M. \& Phillips, T. G. 350 micron observations of local luminous infrared galaxies and the temperature dependence of the emissivity index. \textit{Astrophys. J.} \textbf{662}, 284-293 (2007).
\bibitem{CF} Markevitch, M. \& Vikhlinin, A., Shocks and cold fronts in galaxy clusters. \textit{Phys. Rept.} \textbf{443}, 1-53 (2007).
\bibitem{Old} Howes, M., Casey, A. R., Asplund, M., et al. Extremely metal-poor stars from the cosmic dawn in the bulge of the Milky Way. \textit{Nature} \textbf{527}, 484 - 486 (2015).
\bibitem{Ellis} Hashimoto, T., Laporte, N., Mawatari, K., Inoue, A. K., eet al. The onset of star formation 250 million years after the Big Bang. \textit{Nature} \textbf{557}, 392-394 (2018).
\bibitem{z10} Bouwens, R. J., Illingworth, G. D., Labbe, I., et al. A candidate redshift $z\sim 10$ galaxy and rapid changes in that population at an age of 500 Myr. \textit{Nature} \textbf{469}, 504-506 (2011).
\bibitem{SB} Madau, P. \& Dickinson, M. Cosmic Star Formation History. \textit{Annu. Rev. Astron. Astrophys.} \textbf{52}, 415-491 (2014).
\bibitem{zR} Daddi, E., Renzini, A., Pirzkal, N., et al. Passively evolving early-type galaxies at $1.4<z<2.5$ in the Hubble Ultra Deep Field. \textit{Astrophys. J.} \textbf{626}, 680 - 697 (2005).
\bibitem{LIG} Genzel R., Lutz D., Sturm E., et al. What powers ultra-luminous IRAS galaxies? \textit{Astrophys. J.} \textbf{498}, 579 (1998).
\end{thebibliography}
\end{document}